\documentstyle[aps,prl,twocolumn,psfig]{revtex}

\draft
\begin{document}
\newcommand {\be}{\begin{equation}}
\newcommand {\ee}{\end{equation}}
\newcommand {\bea}{\begin{eqnarray}}
\newcommand {\eea}{\end{eqnarray}}
\newcommand {\nn}{\nonumber}

\twocolumn[\hsize\textwidth\columnwidth\hsize\csname@twocolumnfalse%
\endcsname

\title{Field-Induced Magnetic Order in 
Quantum Spin Liquids} 

\author{Stefan Wessel, Maxim Olshanii, and Stephan Haas}   
\address{Department of Physics and Astronomy, University of Southern
California, Los Angeles, CA 90089-0484 }

\date{\today}
\maketitle

\begin{abstract}

We study magnetic field-induced three-dimensional ordering transitions
in low-dimensional quantum spin liquids, such as weakly coupled,
antiferromagnetic spin-1/2 Heisenberg dimers and ladders. 
Using stochastic series expansion quantum Monte Carlo simulations, 
thermodynamic response functions are obtained down to ultra-low 
temperatures.  
We extract the critical scaling exponents which dictate the 
power-law dependence of the transition temperature on the applied 
magnetic field.
These are compared with recent experiments
on candidate materials and with predictions for the Bose-Einstein 
condensation of magnons obtained in mean-field theory.

\end{abstract}
]

Many low-dimensional quantum spin liquids, such as 
antiferromagnetic Heisenberg spin ladders, have 
a non-magnetic valence-bond ground state with
a finite energy gap 
to the lowest band of triplet excitations. This spin gap 
can be suppressed by an applied magnetic field. 
Increasing the external
field beyond a critical value $h_{c1}$ leads to 
partial spin-polarization and incommensurate,
gapless excitations. A second transition occurs at a higher 
critical field, $h_{c2}$, above which the system becomes fully 
polarized.
It has been suggested that 
in the partially polarized phase, $h_{c1} < h < h_{c2}$,
weak couplings between the low-dimensional 
subsystems induce
a three-dimensional (3D) condensation of magnetic excitations
at low temperatures.\cite{giamarchi,wessel}

Indications of such 3D ordering transitions induced by external
magnetic fields were observed by high-field nuclear magnetic resonance
and inelastic neutron scattering measurements on the quantum spin 
liquids $\rm{TlCuCl_3}$ \cite{oosawa,tanaka} and 
$\rm{Cu_2(C_5H_{12}N_2)_2Cl_4}$ \cite{chaboussant,calemczuk,hammar,mayaffre}. 
The crystal structure of $\rm{TlCuCl_3}$
suggests that this compound consists of weakly coupled 
$\rm{Cu_2Cl_6}$ dimers, involving two neighboring
spin-1/2 $\rm{Cu^{2+}}$ ions in its
b-c plane.\cite{willett,takatsu}  A small spin gap of $\Delta \approx
7.5$K was determined from measurements of the magnetic susceptibility. 
\cite{oosawa,shiramura}
On the other hand, $\rm{Cu_2(C_5H_{12}N_2)_2Cl_4}$ is generally viewed as 
a compound of weakly coupled two-leg ladders with intra-ladder exchange 
constants $J_{\perp} \approx 13.2$K, $J_{\parallel} \approx 2.5$K, and 
a spin gap $\Delta \approx 10.5$K. \cite{chaboussant}
However, a very recent study has concluded that this material consists 
of frustrated planar networks of dimers.\cite{broholm} 
We may further expect field-induced 3D ordering phenomena to occur in 
$\rm (C_5H_{12}N)_2CuBr_4$.\cite{patyal} This compound is believed to 
consists of weakly coupled two-leg ladders with a strong rung coupling,
$J_{\perp} \simeq 13.3$K, between adjacent $\rm CuBr_4$ tetrahedra,
and a weaker coupling, $J_{\parallel} \simeq 3.8$K along the legs
of the ladders.\cite{watson} Its spin gap was observed to be $\Delta \approx
9.5$K.   
 
Whatever the effective dimensionalities of these compounds may turn out to 
be (d=0 for $\rm{TlCuCl_3}$, d=1 or d=2 for $\rm{Cu_2(C_5H_{12}N_2)_2Cl_4}$, 
and d=1 for $\rm (C_5H_{12}N)_2CuBr_4$), they share the essential
common feature of a spin-liquid ground state and
a small but finite spin gap that can be overcome by  
presently accessible magnetic fields.   

From the theoretical side, it has been proposed that the observed 
field-induced 3D magnetic ordering transition in these systems
can be interpreted as a Bose-Einstein condensation of magnons, resulting in a staggered, transverse magnetic order.\cite{giamarchi,nikuni,nikunishiba,honecker}

 This class of quantum phase transition is characterized by a critical 
exponent $\alpha$, which relates the ordering temperature $T_c$ to 
the applied magnetic field according to
\bea
T_c(h) \propto |h - h_c|^{1/ \alpha}.
\eea
The exact determination of this power-law dependence
from experimental data turns out to be 
somewhat delicate, because it requires simultaneous 
fitting of the critical field and the scaling exponent. 
Furthermore, the number of data points tends to be sparse in the
scaling regime around $(h_c , T_c)$. Nevertheless, the 
experimentally deduced exponents, $\alpha \approx 2.0$ for
$\rm{TlCuCl_3}$ and $\alpha \approx 1.5 $ for $\rm{Cu_2(C_5H_{12}N_2)_2Cl_4}$, 
appear to be within the general range of the value $\alpha_{IBEC} = 3/2$, 
expected from isotropic Bose-Einstein Hartree-Fock 
theory. \cite{giamarchi,nikuni,nikunishiba}  
However, material-specific details such as spin-phonon coupling and 
magnetic frustration in these compounds are likely to influence  
the precise values of the scaling exponents. 
 
In this paper, we address the fundamental question of obtaining these 
critical scaling properties directly from microscopic models of weakly 
coupled low-dimensional quantum spin systems. We apply the recently 
developed stochastic series expansion quantum Monte Carlo (QMC)
technique\cite{sandvik} to the 3D antiferromagnetic spin-1/2 Heisenberg 
model with spatially anisotropic exchange couplings, 
\bea
H = \sum_{\langle i,j \rangle} J_{ij} {\bf S}_i \cdot {\bf S }_j
- h \sum_i S^z_i , 
\eea
where $h$ denotes the applied magnetic field. 
The relative strengths of the nearest-neighbor exchange coupling
constants $J_{ij}$ are illustrated in Fig.~1, in which planar sections 
of two clusters with different configurations of $J_{ij}$ are shown.
Fig.~1(a) illustrates an ensemble of weakly coupled dimers oriented 
along the x-direction, a configuration resembling the minimal effective 
magnetically structure of $\rm{TlCuCl_3}$. Fig.~1(b) shows a 
quasi-1D array of weakly coupled two-leg Heisenberg 
ladders oriented along the y-direction, as may be realized in 
$\rm (C_5H_{12}N)_2CuBr_4$ or $\rm{Cu_2(C_5H_{12}N_2)_2Cl_4}$.

\begin{figure}[h]
\centerline{\psfig{figure=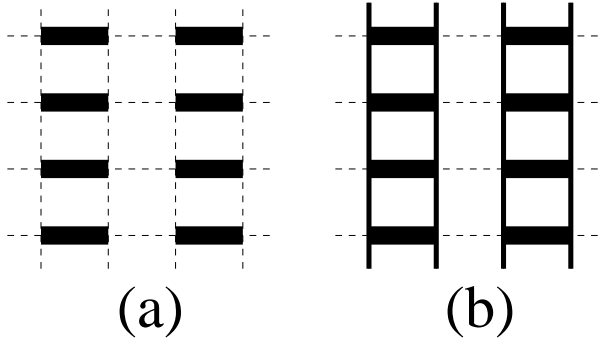,width=8cm,angle=0}}
\vspace{0.3cm}
\caption{
Layers of anisotropically coupled quantum spins:
(a) weakly coupled dimers, and (b) weakly coupled ladders.
The strength of the exchange coupling constants is indicated
by the thickness of the lines. The 3D
inter-layer couplings (not shown in the figure) 
have the same strength as the weakest links (dashed lines)
within the planes.
}
\end{figure}

The numerical algorithm
involves expansions of the partition function in inverse temperature,
uses local and global system updates, and is significantly more
efficient than conventional QMC schemes. In this work we study
cubic lattices with up to $10 \times 10 \times 10$ quantum spins
down to very low temperatures, by which is meant less than 1\% of 
the exchange coupling strength $J$ which sets the scale of the problem. 
Moreover, the stochastic series expansion QMC method
can handle external magnetic fields of any strength
without the problems common to 
worldline QMC techniques, such as the QMC loop algorithm. 

In Fig.~2(a), the low-temperature regime of the uniform magnetization 
is shown for the system of weakly coupled dimers depicted in Fig.~1(a)
at various magnetic fields within the partially polarized regime,
$h_{c1} < h < h_{c2}$.
The anisotropy of the exchange coupling constants, $J'/J = 1/15$, was 
chosen according to estimates for $\rm{TlCuCl_3}$. Here 
$J$ is the strong intra-dimer and $J'$ the weak inter-dimer 
coupling. In this case, 
the observed critical fields are well approximated by
perturbation theory about the limit of non-interacting dimers,
giving $h_{c1} \simeq J - 5 J'/2$ and $h_{c2} \simeq J + 5 J'$. 
At ultra-low temperatures on the order of $J'$, the magnetization 
curves of the weakly coupled dimers (solid lines) have a maximum if 
$h_m < h < h_{c2}$ and a minimum if $h_{c1} < h < h_m$, where 
$h_m \equiv (h_{c1}+h_{c2})/2$, indicating the onset of a magnetic 
field-induced 3D ordering. This feature, emphasized by large
filled circles, is absent in the magnetization curves for the 
non-interacting limit ($J' = 0$), denoted by the dashed lines. 
While effects of the weak inter-dimer couplings are clearly negligible 
at higher temperatures, their relevance at low temperatures is 
seen in the departure of the solid lines from the dashed lines. 
In contrast to the mean-field theory of weakly coupled dimers,
\cite{tachiki}
the magnetization is observed to be strongly temperature-dependent below the 
ordering temperature $T_c$, as shown in Figs.~2(b) and (c). 
On the other hand, the 3D
behavior is in good qualitative agreement with the recently proposed
Bose-Einstein condensation (BEC) description of this ordering
transition.\cite{nikuni} 

\begin{figure}[h]
\centerline{\psfig{figure=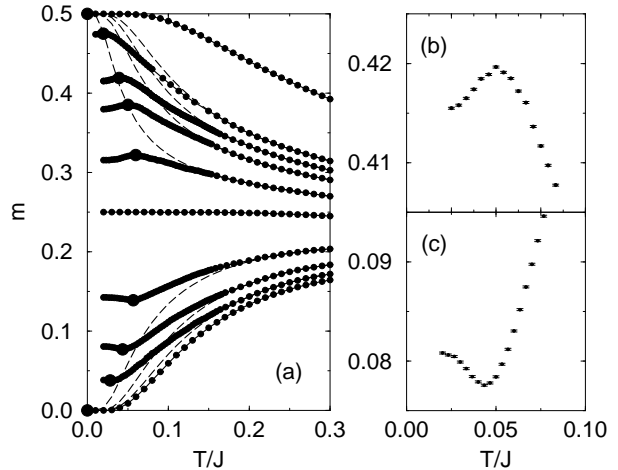,width=8cm,angle=270}}
\vspace{0.3cm}
\caption{(a)
Temperature dependence of the uniform magnetization in the system of
weakly coupled dimers shown in Fig.~1(a). 
The coupling anisotropy is $J'/J = 1/15$, and the magnetic fields
are $h/J$ = 0.80, 0.86, 0.90, 0.97, 1.08, 1.17, 1.23, 1.27, 1.32,
1.37. Extrema in $m(T)$ indicate the onset of 
3D ordering, and are denoted by enlarged filled circles. Results for the 
non-interacting ($J'/J = 0$) limit are plotted with dashed lines. 
(b) Low-temperature regime of $m(T)$ at $h = 1.27 J$ and (c) at 
$h = 0.9 J$.
}
\end{figure}

The dependence of $T_c$ on the applied magnetic field can be 
extracted from the locus of the extrema in $m(T)$ which are robust to 
finite size effects for the system sizes considered here.\cite{footnote1}
The resulting phase diagram is plotted in Fig.~3 for the ensemble of 
weakly coupled dimers. As expected from Eq.~1, the transition temperature
exhibits a power-law dependence in the vicinity of
$h_{c1}$ and $h_{c2}$. Our best fits, shown in the insets of Fig. 3, 
yield  $\alpha = 2.7 \pm 0.2$ for the lower-field transition and
$\alpha = 2.3 \pm 0.2$ in the vicinity of $h_{c2}$. 
These critical exponents 
differ significantly from the value obtained by standard, isotropic
Bose-Einstein Hartree-Fock theory, $\alpha_{IBEC} = 3/2$.   
It was previously pointed out that some of this deviation may be attributed 
to shortcomings of the Hartree-Fock description in the critical regions.
\cite{nikuni}

Another important point is that the
dispersion of the spin triplet band is 
strongly anisotropic.\cite{cavadini} For the case of linearly aligned, 
weakly coupled dimers the dominant feature in the triplet excitation 
spectrum is a parabolic bonding band along the strong-coupling
direction which becomes populated when the magnetic field is raised
beyond $h_{c1}$.\cite{affleck}  Along the weak-coupling directions the triplet excitation spectrum is linear in
this regime. Following the BEC Hartree-Fock treatment of \cite{nikuni} , a critical  exponent $\alpha=5/2$ is obtained for this case.\cite{olshanii}  
This anisotropic BEC Hartree-Fock
exponent  is in better agreement with the QMC 
simulations than $\alpha_{IBEC} = 3/2$.

\begin{figure}[h]
\centerline{\psfig{figure=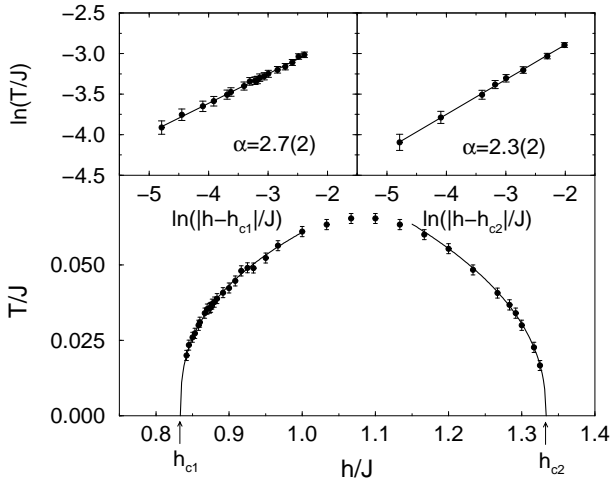,width=8cm,angle=270}}
\vspace{0.3cm}
\caption{Phase diagram of the magnetic field-induced
ordering transition in a system of weakly coupled dimers. 
The scaling exponents in the vicinity of the lower and upper critical fields
are extracted from double-logarithmic plots of 
$T_c(h)$, shown in the insets. 
}
\end{figure}
  
On the other hand, measurements of the lower-field critical exponent
in $\rm{TlCuCl_3}$ give $\alpha \simeq 2.0$\cite{tanaka} and 
$\alpha \simeq 2.2$\cite{nikuni}. This suggests that the generic 
Heisenberg Hamiltonian which we have studied is not sufficient 
to accurately model the critical properties of this compound. 
Reasons for this discrepancy may be found in (i) the much more complex 
bandstructure of the spin-triplet excitations in the real material, 
\cite{cavadini,oosawakato} (ii) magnetic frustration effects, and (iii) 
coupling to phonon degrees of freedom. 

Let us now turn to the system of weakly coupled two-leg
ladders depicted in Fig.~1(b). For the coupling constants 
chosen in Fig. 4 and 5, the critical magnetic fields are
$h_{c1} = 0.61 J$ and $h_{c2} = 1.86 J$. It has recently been pointed 
out that this system exhibits extrema in $m(T)$ even in the complete
absence of inter-ladder couplings.\cite{wang} We can confirm 
this observation with our simulations of the uniform magnetization,  
as shown in Figs.~4(a) and (c). At magnetic fields slightly above 
$h_{c1}$  minima occur at low temperatures, whereas 
maxima are observed at larger magnetic fields close to $h_{c2}$. 
These features indicate the transition into low-temperature Luttinger
Liquid behavior in the partially polarized regime, $h_{c1} < h < h_{c2}$.\cite{wang}
When the weak inter-ladder couplings are switched on, a second feature, 
marked by the filled arrows in Figs.~4(b) and (d), occurs at still lower
temperature. This is the 3D magnetic ordering transition, also
manifested as a change of slope in the energy, as shown in the insets.  
Measurements of the magnetic specific heat in 
$\rm{Cu_2(C_5H_{12}N_2)_2Cl_4}$ indeed show such 3D ordering 
features at temperatures below the onset of quantum critical 
behavior.\cite{calemczuk}

\begin{figure}[h]
\centerline{\psfig{figure=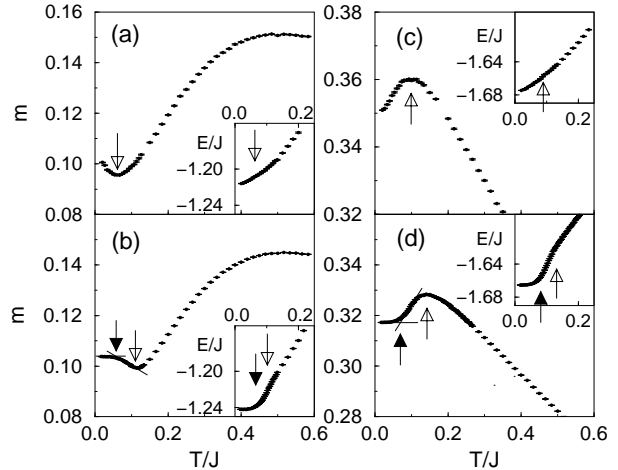,width=8cm,angle=270}}
\vspace{0.3cm}
\caption{Low-temperature regime of the uniform magnetization in
ladder systems. (a) and (c): isolated two-leg ladders. (b) and (d):
system of weakly coupled ladders shown in Fig.~1(b).
The coupling constants are $J \equiv J_{\perp}$, 
$J_{\parallel} = J /3$, $J' = 0$ for (a) and (c),
and $J' = J / 15$ for (b) and (d).
The applied magnetic fields are $h = 0.83 J$ for
(a) and (b), and $h = 1.5 J$ for (c) and (d).
The temperature dependence of the energy is shown in the insets. 
}
\end{figure}

A rich magnetic phase diagram, shown in Fig.~5,
can be constructed from these magnetization response functions,
and shares some of the essential features reported by experiments on 
$\rm (C_5H_{12}N)_2CuBr_4$ and
$\rm{Cu_2(C_5H_{12}N_2)_2Cl_4}$. At high temperatures (not shown here),
the system is effectively zero-dimensional, with a paramagnetic 
Curie-law temperature dependence of the magnetic susceptibility. 
Lowering the temperature, there is a low-field disordered spin 
liquid regime ($h < h_m$) and a high-field spin-polarized phase
($h > h_m$), where $h_m \approx (h_{c1} + h_{c2})/2$ separates these
two regions. The onset of Luttinger Liquid behavior is found 
in the partially spin-polarized regime, $h_{c1} < h < h_{c2}$, 
at temperatures below $0.13 J$ for $J_{\parallel} = J /3$. The 
effective dimensionality of this region is d=1. Finally, 3D 
ordering occurs at still lower temperatures, as indicated by the 
lowest transition line in Fig. 5. The scaling exponents extracted 
from fits of $T_c(h_c)$ to the power law in Eq. 1 are $\alpha = 
3.1 \pm 0.2$ at the lower critical field and $\alpha = 1.8 \pm 0.2$ 
at the upper critical field. These values are quite different from 
isotropic Bose-Einstein Hartree-Fock theory, suggesting that a 
successful effective theory needs to account more accurately for 
the quantum dynamics of the lower-dimensional subsystems.  
 
\begin{figure}[h]
\centerline{\psfig{figure=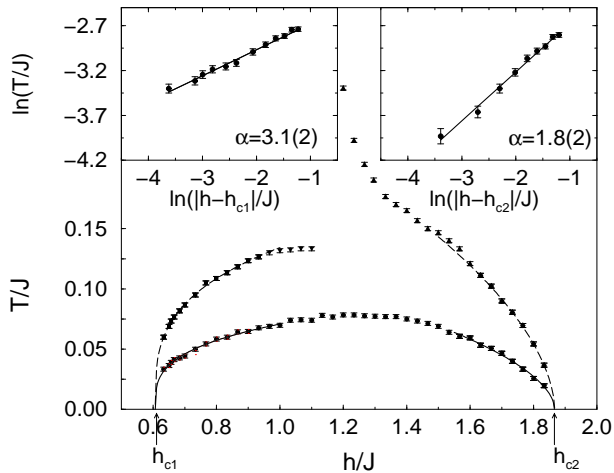,width=8cm,angle=270}}
\vspace{0.3cm}
\caption{Magnetic phase diagram of weakly coupled ladders.  
The upper lines indicate the transition into a Luttinger 
Liquid regime, also present in isolated two-leg ladders.  
The lower line marks the magnetic field induced 3D
ordering transition. 
The scaling exponents in the vicinity of the lower and upper critical fields
are extracted from double-logarithmic plots of 
$T_c(h)$, shown in the insets. 
}
\end{figure}

In summary, we have studied the magnetic phase diagram and the critical 
behavior in models of weakly 
coupled quantum spin liquids. Using the stochastic series expansion QMC
method, we were able to reach sufficiently low temperatures that magnetic
field-induced 3D ordering could be observed. The scaling 
exponents depend strongly on the dimensionality and on the 
quantum dynamics of the subsystems, Heisenberg dimers and ladders,
which reflect the strongly anisotropic dispersion of the triplet excitation 
bands. While this study has concentrated on the fundamental features of
generic model Hamiltonians, much more detailed models are clearly needed to 
account for the specific scaling properties observed in real materials.     

We thank 
N. Cavadini, A. Honecker, B. Normand, M. Oshikawa,  and G. S. Uhrig
for useful discussions,
and acknowledge financial support by the National Science Foundation,
Grant No. DMR-0089882.


\begin{references}
\bibitem{footnote1} In practice, we monitor the numerical 
derivative of $m(T)$ to obtain a more precise location of
$T_c(h)$.
\end{references}
\end{document}